# Prediction of Local Failure after Stereotactic Radiotherapy in Melanoma Brain Metastases Using Ensemble Learning on Clinical, Dosimetric, and Radiomic Data

Nanna E. Wielenberg[1,7], Ilias Sachpazidis[2,7], Oliver Blanck[3], Lucas Zander[4], Jan C. Peeken[4], Stephanie E. Combs[4], Horst Urbach[5], Maxim Zaitsev[6,7], Dimos Baltas[2,7], Ilinca Popp[1,7], Anca-Ligia Grosu[1,7], Tobias Fechter[2,7]

[1] Department of Radiation Oncology, Medical Center University of Freiburg, Medical Faculty University of Freiburg, Freiburg, Germany

[2] Division of Medical Physics, Department of Radiation Oncology, Medical Center University of Freiburg, Medical Faculty University of Freiburg, Freiburg, Germany

[3] Department of Radiation Oncology, University Medical Center Schleswig-Holstein, Kiel, Germany

[4] Department of Radiation Oncology, Klinikum Rechts der Isar, Technical University of Munich (TUM), Munich, Germany

[5] Department of Neuroradiology, Medical Center University of Freiburg, Medical Faculty University of Freiburg, Freiburg, Germany

[6] Division of Medical Physics, Department of Diagnostic and Interventional Radiology, Medical Center University of Freiburg, Medical Faculty University of Freiburg, Freiburg, Germany

[7] German Cancer Consortium (DKTK). Partner Site Freiburg, Freiburg, Germany

**Corresponding Author:**

Nanna E. Wielenberg
Department of Radiation Oncology
Medical Center – University of Freiburg
Robert-Koch-Str. 3
79106 Freiburg im Breisgau, Germany
E-mail: nanna.wielenberg@uniklinik-freiburg.de

**Abstract**

Background: This study aimed to predict lesion-specific outcomes after stereotactic radiotherapy (SRT) in patients with brain metastases from malignant melanoma (MBM), using clinical, dosimetric, and pretherapeutic MRI data.

Methods: In this multicenter retrospective study, 517 MBM from 130 patients treated with single-fraction or hypofractionated SRT at three centers were analyzed. From contrast-enhanced T1-weighted MRI, 1576 radiomic features (RF) were extracted per lesion - 788 from the gross tumor volume (GTV) and 788 from a 3 mm peritumoral margin. Clinical, dosimetric and RF data from one center were used for feature selection and model development via nested cross-validation employing an ensemble learning approach; external validation used data from the other two centers.

Results: Local failure occurred in 72/517 lesions (13.9%). Predictive models based on clinical data, RF, or a combination of both achieved c-indices of 0.60 ± 0.15, 0.65 ± 0.11, and 0.65 ± 0.12, respectively. RF-based models outperformed the clinical models; dosimetric data alone were not predictive. Most predictive RF originated from the peritumoral margin (92%) versus GTV (76%). On the first external dataset, all models performed similarly (c-index: 0.60–0.63), but generalization was poor on the second (c-index < 0.50), likely due to differences in patient characteristics and imaging protocols.

Conclusions: Pretherapeutic MRI features, particularly from the peritumoral region, show promise for predicting lesion-specific outcomes in MBM after SRT. Their consistent contribution suggests biologically relevant information that may support individualized treatment planning. Combined with clinical data, these markers offer prognostic insight, though generalizability remains limited by data heterogeneity.

**Abbreviations:**

**BED –** Biologically Equivalent Dose
**CPHM –** Multivariate Cox Proportional Hazard Models
**GTV –** Gross Tumor Volume
**KPS –** Karnofsky Performance Scale
**LF –** Local Failure
**MBM** – Malignant Melanoma
**RF** – Radiomic Features
**SFRT** – Stereotactic Fractionated Radiotherapy
**SRS** – Stereotactic Radiosurgery
**SRT** – Stereotactic Radiotherapy

## 1.0 INTRODUCTION

Malignant melanoma is a leading cause of brain metastases. Nearly 50% of patients with stage IV disease develop brain metastases[1]. Advances in imaging, local treatments, and systemic therapies have improved the historically poor prognosis of melanoma brain metastases (MBM)[2] and increased the long-term disease control[3]. However, MBM remain a major cause of intracranial failure and neurological death, even in patients with controlled extracranial disease[4]. MBM are more likely to cause neurological death than those from other cancers [5], highlighting the importance of effective local therapy[6,7].

Stereotactic radiotherapy (SRT), including radiosurgery (SRS) and stereotactic fractionated radiotherapy (SFRT), plays a key role in treating MBM. With one-year control rates of 60-80%[8,9], SRT offers high survival benefits, particularly for patients with solitary or oligometastatic disease[10]. Even with 10 or more MBM, SRS without whole-brain radiotherapy provides good intracranial control[11]. When whole-brain radiotherapy is needed, hippocampal-sparing techniques can help avoid cognitive deficits[12].

While some lesions respond well to stereotactic radiotherapy, others show local failure (LF), and the underlying factors driving this variability remain poorly understood. Several patient-specific factors, such as lesion number, intracranial tumor volume, and age, have been linked to survival after SRT[13,14], but these do not predict local treatment response. Lesions showing LF may have unique characteristics detectable through analysis of radiomic features (RF). RF analysis extracts quantitative features from clinical imaging and has been suggested as a potential tool for predicting intratumoral heterogeneity and risk of intracranial progression[15,16].

Machine learning holds promise for predicting treatment response and local failure in patients with brain metastases undergoing stereotactic radiotherapy[17]. Integrating advanced machine learning techniques - such as ensemble learning - with radiomic and clinical data offers the potential to improve individualized lesion-level prediction of treatment response. This could enable tailored treatment strategies that improve patient stratification, intracranial control, and overall outcomes.

Our study aims to investigate the predictive value of clinical, dosimetric, and MRI features for lesion-specific outcomes in MBM following SRT, using an ensemble learning approach.

## 2.0 MATERIALS & METHODS

### 2.1 Patient characteristics

In this multicenter, retrospective analysis, we included neuroimaging, clinical and dosimetric data on 517 metastases from 130 patients who received 179 series of SRT for MBM. Inclusion criteria were as follows: (a) local radiotherapy for brain metastases from pathologically confirmed malignant melanoma between 2012 and 2021; (b) age ≥ 18 years; (c) pretherapeutic contrast-enhanced MRI; (d) minimum biologically equivalent dose (BED) of 41 Gy (α/ß=10 Gy). Lesions previously treated with neurosurgery or radiotherapy were excluded. The majority of patients were treated with SRS or SRS combined with whole-brain radiotherapy. In latter case, lesions were treated with 51 Gy in 12 fractions, with hippocampal sparing, according to the HIPPORAD trial protocol[18]. Treatments were administered across three primary care centers. Clinical parameters and radiotherapy specifications evaluated as potential predictors are summarized in Tables 1 and 2, respectively. The dosimetric features considered included minimum, maximum, and mean dose, as well as D50%, D98%, and D2%.

The aim of this study was to identify predictive markers of lesion-specific LF using clinical, dosimetric, and radiomic features extracted from pretherapeutic MRI. The primary outcome was LF, defined per lesion as in-field progression according to the Response Assessment in Neuro-Oncology brain metastases (RANO-BM) criteria[19]. LF was assessed by MRI 6-8 weeks post-treatment and every 3 months thereafter, with a minimum follow-up of 12 weeks.

## 2.2 Image-based features

RF were extracted from Magnetization Prepared Rapid Gradient Echo 3D T1-weighted MRI, acquired across three institutions using different scanners and slighty varying acquisition protocols. Detailed parameters are summarized in Supplementary Table I.

To standardize feature extraction, all MR images underwent the following pre-processing steps: First, bias field correction[20] was applied using ANTs software v2.3.5[21]. The datasets were then resampled to a voxel length of 1 mm in each dimension, and the voxel intensities were adjusted within an automatically generated brain mask (FSL brain extraction tool v6.0[22]) using Z-score normalization.

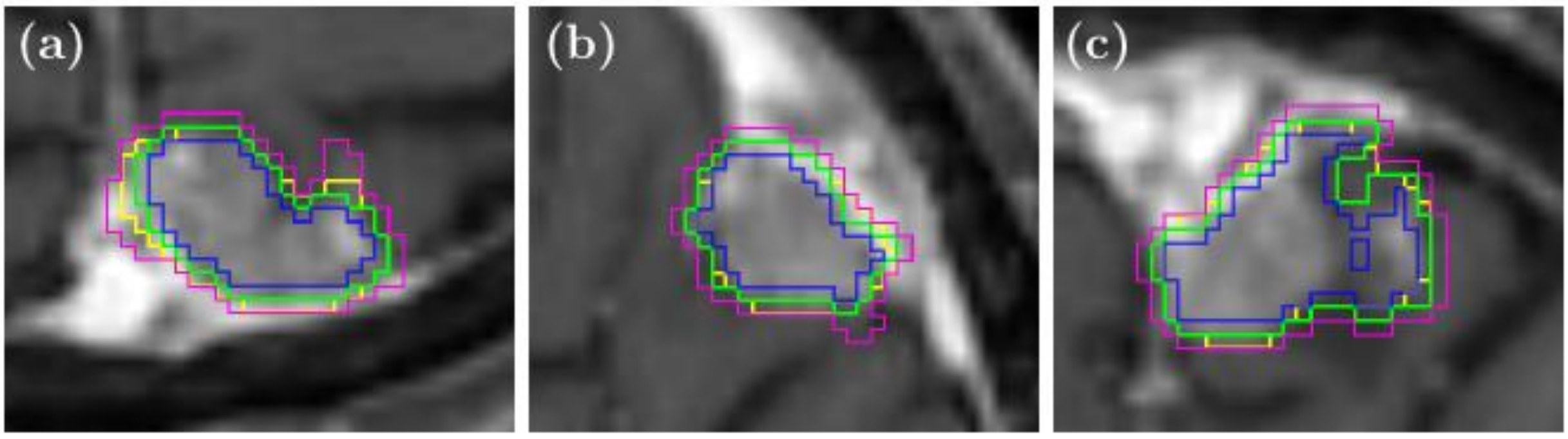

**Figure 1**: Axial (a), sagittal (b) and coronal (c) views of the contours used to simulate inter-observer variability. The expert contour is shown in green, the artificial contour is shown in yellow and the expert contour with erosion or dilation is shown in blue and pink, respectively.

The initial contours, representing the gross tumor volume (GTV) were manually delineated by different radiation oncologists on 3D-GdT1w-MRI used for radiotherapy planning. An expert radiation oncologist reviewed all contours prior to analysis. To address potential inter-observer variability without the time-consuming task of adding more manual contours, three artificial contour datasets were created in addition to the manual GTV contours (Figure 1). One artificial dataset was generated using an in-house trained nnU-Net[23]. The other two datasets were created by applying morphological erosion, dilation, or an identity function to the manual contours. Further details are provided in Supplementary Material I.

RF were calculated for the segmented tumor and a 3 mm isotropic margin around it to account for microscopic infiltration[24] using Pyradiomics v3.0.1[25]. A total of 1576 RF were calculated for each contour, with 788 for the tumor and 788 for the margin.

### 2.3 Feature elimination and model-building

Four model types were trained to estimate lesion-specific LF risk: models based on clinical data, dosimetric parameters, RF and a combined model. The entire processing pipeline is shown in Figure 2. A 30x30 nested cross-validation scheme[26] was used for feature elimination, model building, and evaluation. In an outer loop, the dataset was divided 30 times into a training set and test set (80:20). Each training set is further split in the inner loop (80:20) into a training and validation set. Feature elimination and model building occur on the inner loop's training sets, while the validation sets are used for model ranking. The best models from each inner loop are combined into an ensemble model, which is evaluated on the outer loop's test set. The final performance estimate is the average performance across all 30 outer loop test sets. This methodology ensures independent evaluation of selected models. The datasets were disjoint at the patient level, preventing overlap of lesions from the same patient between sets.

To reduce overfitting, features were filtered on the inner training sets based on variability, redundancy, and correlation with LF. RF were reduced by removement of contour-dependent features, univariate filtering and clustering of correlated features. Details are in Supplementary Material II. Clinical and dosimetric features were filtered via univariate analysis only.

The nested cross-validation estimated the performance of the entire pipeline, not a single model[26]. For external validation, ensemble models from the outer loop were applied to test data, and the mean concordance index (c-index) ± standard deviation was calculated. For each model type, three performance metrics were reported: (1) Mean c-index ± SD from nested cross-validation using data from center I, (2) Mean c-index ± SD for external validation on center II and (3) Mean c-index ± SD for external validation on center III.

Multivariate Cox proportional hazard models (CPHM)[27] were used to model lesion-specific LF risk based on clinical, dosimetric and image features. Multivariate CPHM were constructed as follows: Univariate models were first fitted on the training set using the features remaining after elimination. Starting with the feature yielding the highest c-index on the validation set, additional features were added iteratively if they improved the c-index, up to a maximum of four features per model.

Ensemble models combined individual CPHM by averaging their risk scores after excluding outliers (values beyond ±2 standard deviations). This ensemble approach helps mitigate overfitting, compensates for weaker individual models, and improves robustness

The developed code, RF, and models are publicly available: https://github.com/ToFec/RadiomicsMM.

**2.4 Statistical analysis**

All analyses were conducted in R v4.1.2. Group differences were assessed using the Wilcoxon signed-rank test; survival curves were compared using log-rank tests. A significance level of 5% was used throughout, except for RF contour-dependence tests, where a 10% threshold was applied to reduce feature count.

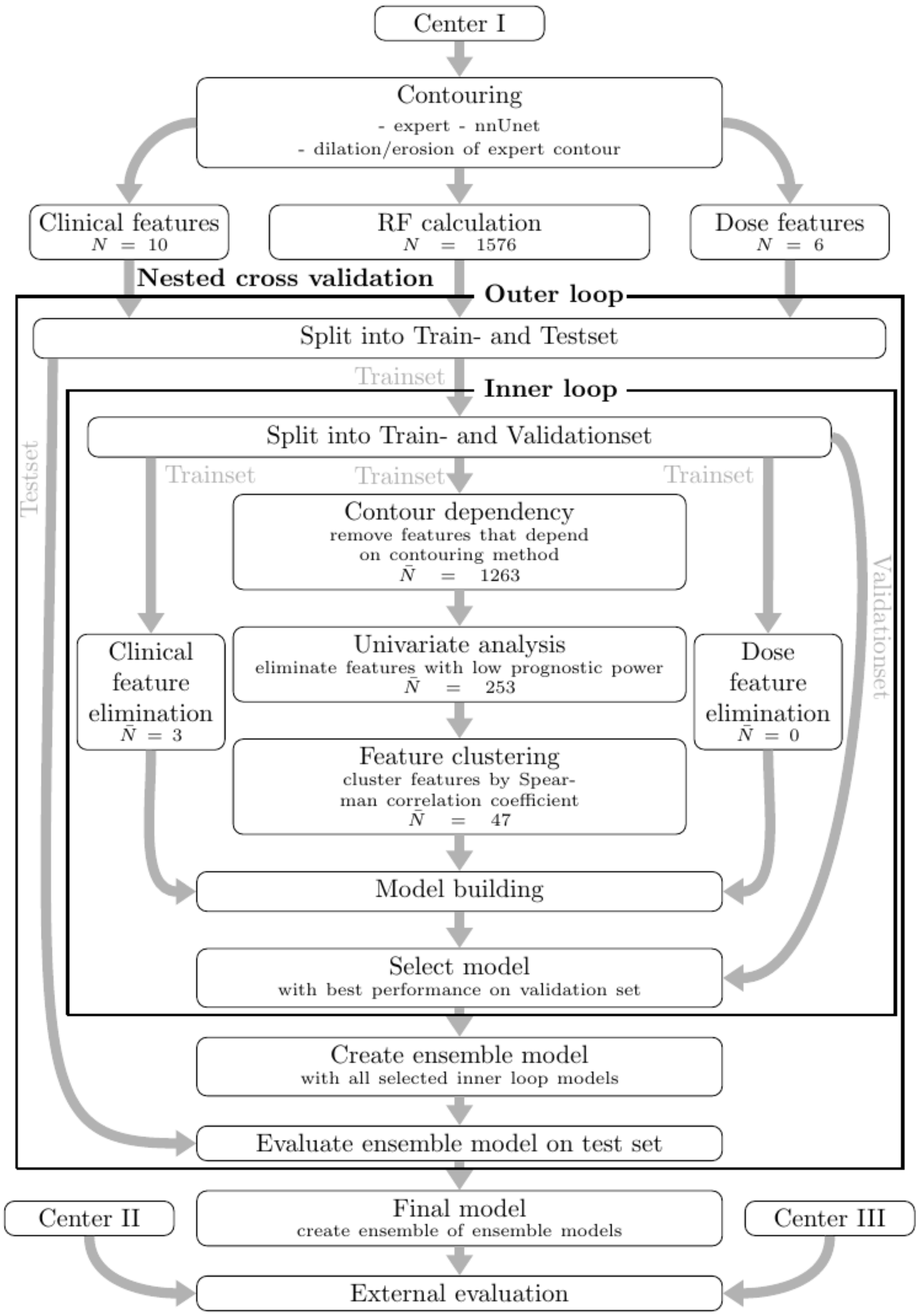

**Figure 2**: In a first step, four contour sets were generated per lesion. Radiomic, dosimetric, and clinical features were then extracted. Feature elimination and model building followed a 30×30 nested cross-validation scheme. In each inner loop, redundant and contour-dependent features were removed, and Cox proportional hazards models were trained and ranked. The top models of each inner loop were combined into an ensemble model and evaluated on the respective outer loop test set. Finally, all ensemble models from the outer loop were applied to two external test sets for validation.

## 3.0 RESULTS

### 3.1 Clinical characteristics

The mean age of patients was 61.9 ± 14.2 years across all centers, with comparable age distributions throughout the cohorts. The median number of MBM at the start of SRT was 3 (range 1–30). A substantial proportion of patients in Center I (41/83) and Center II (25/77) presented ≥4 BMs, whereas only 4 out of 19 patients in Center III had more than two BMs.

Karnofsky performance scale (KPS) was highest in Centers I and II (66/83 and 61/77 with KPS 90–100%), whereas most patients in Center III had lower scores (only 7/19 with KPS 90–100%). Systemic therapy use was most frequent in Center II (before and during RT) and lowest in Center III prior to RT, though post-RT treatment was relatively common. Melanoma Molecular-Graded Prognostic Assessment scores indicated more favorable prognoses in Center II, with Center I having more patients in lower score ranges; data were unavailable for Center III.

SRS was the main treatment modality (349/517 lesions), most commonly used in Center II (218/226 lesions), followed by Centers I and III. The median BED was 180 Gy (α/β = 2 Gy) and 60 Gy (α/β = 10 Gy), with no significant differences between lesions with and without LF. BED distributions were similar across centers, with Center II showing slightly lower median BED for α/β = 2 Gy and higher for α/β = 10 Gy, indicating less frequent use of fractionated regimens. D2% and D98% values were comparable between lesions with and without LF.

LF occurred in 13.9% of lesions (72/517) at a mean of 6.6 months. At the series level, LF was observed in 22.3% (40/179), with 25.4% of patients (33/130) affected. The 12-month local control rate was 88.6%, median survival 16.3 months, and LF-free survival 13.4 months. LF rates were lowest in Center II (6.2%) and highest in Center III (22.7%).

### 3.2 Model-building and analysis

We evaluated the processing pipeline using four model types: clinical, dosimetric, radiomic, and a combined model. External validation was performed on two independent datasets.

In models based solely on clinical features, the KPS, systemic therapy more than 14 days before SRT, and systemic therapy during SRT were most predictive of local failure. These variables were included in 68%, 43%, and 35% of the selected models, respectively, with corresponding mean hazard ratios (HR ± SD) of 0.51 ± 0.06 (KPS), 0.41 ± 0.06 (systemic therapy before SRT), and 0.39 ± 0.04 (systemic therapy during SRT).

The clinical-only model achieved a mean c-index of 0.60 ± 0.15 in internal validation (center I), and 0.60 ± 0.01 and 0.47 ± 0.06 on the external datasets from centers II and III, respectively. Kaplan-Meier curves (Figure 3) showed limited separation of high- and low-risk groups in centers II and III, especially in late events. However, survival differences between risk groups were statistically significant in all datasets.Model building was successful in 5% of the iterations when only dosimetric features were used as input to the processing pipeline. In the remaining 95%, no feature remained after elimination. Therefore, dosimetric features were not further analyzed.

When the pipeline was fed solely with RF, the mean c-index was 0.65 ± 0.11 for center I. Performance decreased slightly for center II (c-index: 0.62 ± 0.02), but dropped significantly for center III (c-index: 0.46 ± 0.03). This drop is reflected in the survival curves in Figure 4, where high-risk patients had better LF-free survival than low-risk patients. Statistical analysis confirmed significant survival differences between high- and low-risk patients across all datasets.

Notably, most features used in the selected models were derived from the tumor margin. Features from the 3 mm margin were included in 92% of models, compared to 76% for GTV features. The most prominent RF, present in 22% of models, was *Margin_wavelet.LLL_firstorder_90Percentile*, with a mean HR of 0.34 ± 0.07. This feature was extracted from the tumor margin, and "*wavelet.LLL*" indicates a low-pass filter was applied to the image along all axes.

Combining clinical data with radiomic features improved both the performance and generalizability of the models. Compared to the pure radiomics model, the average c-index remained stable in Center I (0.65 ± 0.12) but increased slightly in Centers II (0.63 ± 0.02) and III (0.48 ± 0.03). Kaplan–Meier curves (Figure 5) demonstrate improved separation between high- and low-risk groups in Center II, compared to the radiomics-only (Figure 4) and clinical-only (Figure 3) models. Despite this, overall performance in Center III remained low. The most influential features mirrored those identified in the mentioned clinical and radiomics models. Survival differences between high- and low-risk groups were statistically significant in all centers except Center III.

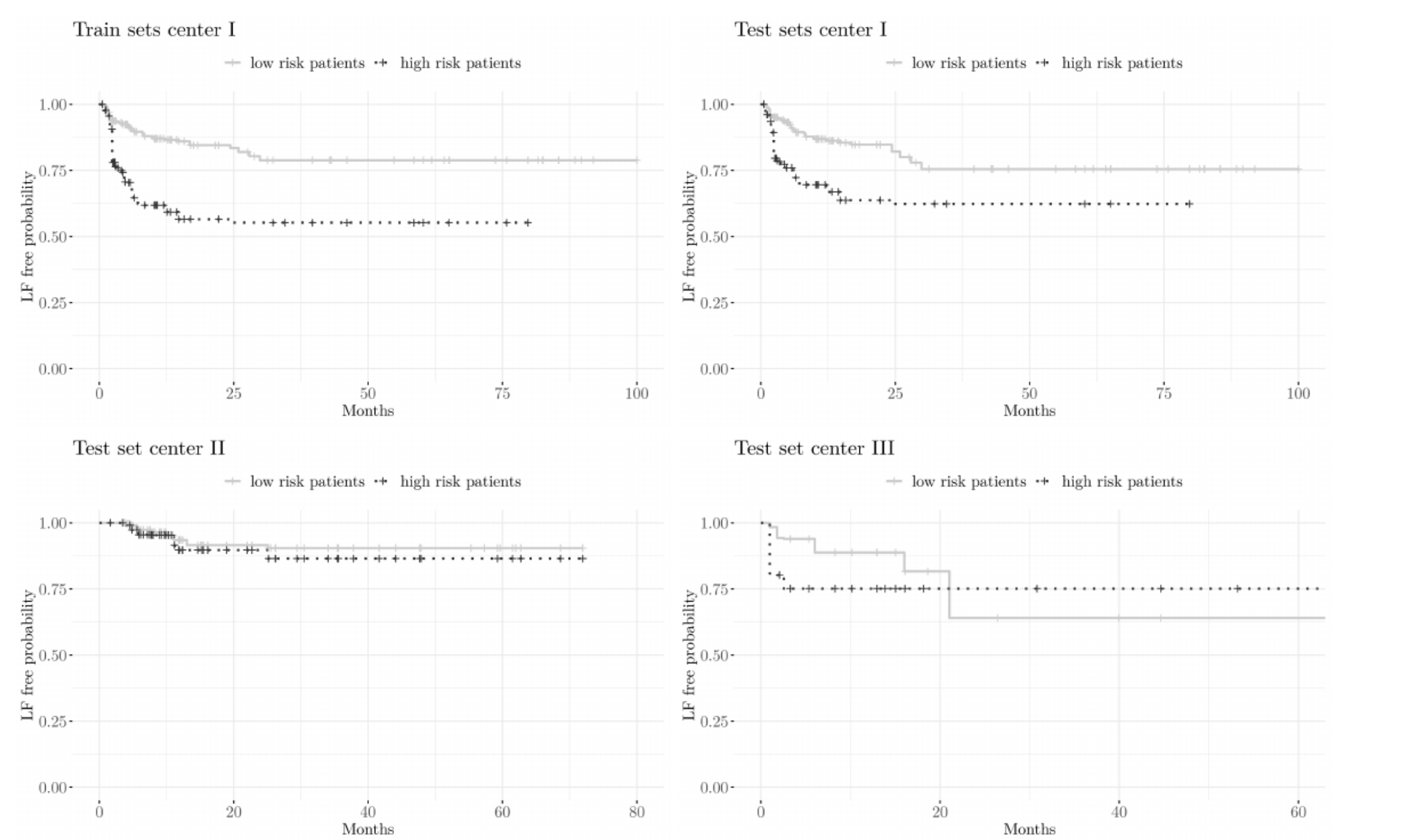

**Figure 3**: Kaplan-Meier plots showing LF-free probabilities for high- vs. low-risk patients based on models using only clinical features: training cohort (upper left) and test sets from center I (upper right), center II (lower left), and center III (lower right).

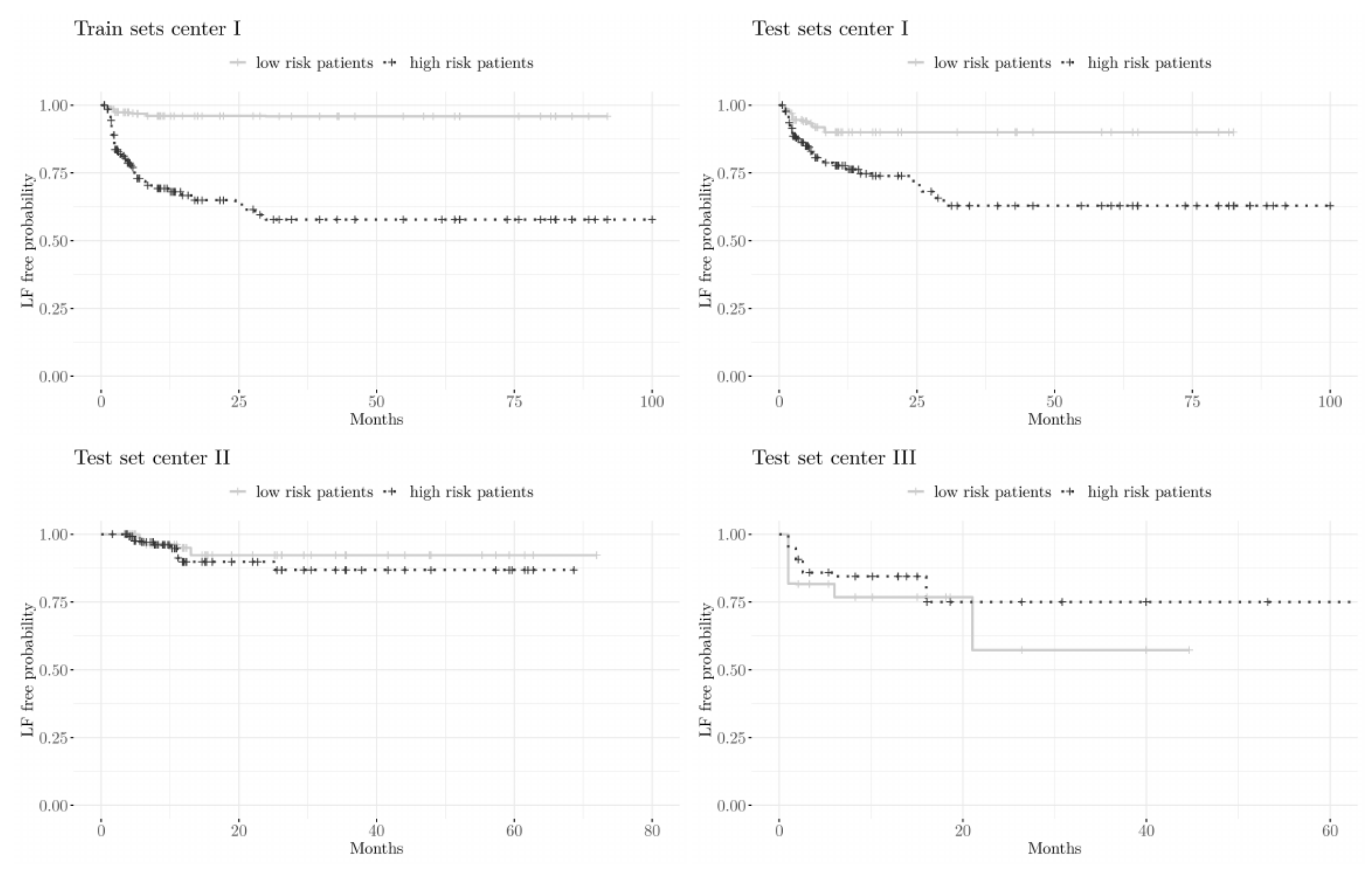

**Figure 4**: Kaplan–Meier plots of LF-free probabilities for high- vs. low-risk patients based on models using only radiomic features: training cohort (upper left) and test sets from centers I–III (upper right to lower right).

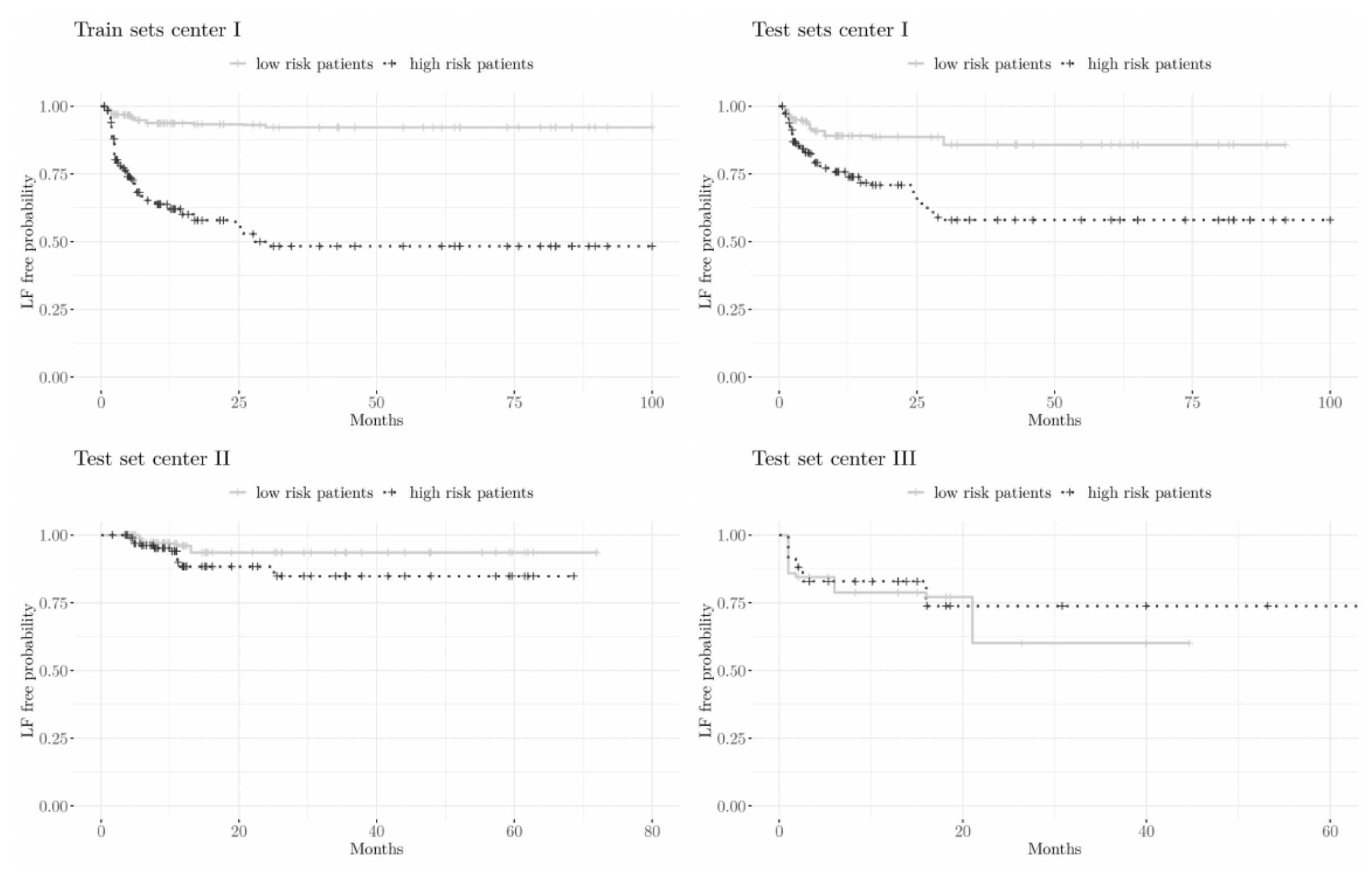

**Figure 5**: Kaplan-Meier plots of LF-free probabilities for high- vs, low-risk patients based on models using both radiomic and clinical features: training cohort (upper left) and test sets from centers I–III (upper right to lower right).

## 4.0 DISCUSSION

This study hightlights the potential of combining clinical and imaging data from pretherapeutic MRI to predict lesion-specific outcomes in patients with MBM undergoing SRT. To our knowledge, this is the first multicenter study presenting predictive imaging markers for SRT of MBM in a large cohort. The use of an ensemble learning approach enabled robust feature-integration and may serve as a framework for future radiomic modeling in heterogeneous, multicenter settings.

Our models achieved c-indices above 0.63 on independent cohorts from two centers, aligning with published literature[28]. While RF alone showed good performance, combining RF with clinical parameters - particularly KPS and systemic therapy information - yielded the best results. These clinical variables are established prognostic markers of survival in MBM, and systemic therapy administered before or during SRT has been associated with improved outcomes[29]. In contrast, dosimetric parameters showed limited predictive value, likely due to similar treatment standards across centers and minimal differences in BED between lesions with and without local failure.

A key finding of this study was the importance of peritumoral margin features. Approximately 92% of the models relied on RF extracted from the margin zone, suggesting that this region provides essential information on recurrence risk. These features may capture subtle peritumoral changes linked to aggressive or infiltrative tumor behavior that conventional MRI cannot detect and that are often excluded from the planning target volume. This aligns with studies using 5-ALA fluorescence, which revealed infiltrative growth beyond imaging-visible tumor margins[30–32]. Radiomics may therefore offer a window into the cellular and molecular microenvironment of brain metastases[33,34]. Previous work has suggested that RF can act as biomarkers of intratumoral heterogeneity and risk of intracranial progression [16]. However, accurate interpretation of RF remains challenging, with a lack of understanding hindering clinical implementation. Correlation with histopathology may help clarify the biological underpinning of these features. Based on our findings, we recommend incorporating peritumoral in future radiomic models predicting LF.

Despite the potential of these models, their generalizability remains a major challenge. While performance remained stable across Centers I and II, it dropped markedly for Center III (mean c-index < 0.50). Predictive features from the training set were either non-informative or even inversely associated with outcome in Center III. This discrepancy likely stems from multiple factors.

First, differences in acquisition matrices played a significant role. Prior studies have shown that matrix parameters influence radiomic features[35], which was confirmed in our dataset through high classification accuracy based on matrix type. Despite efforts to eliminate matrix-dependent features and retrain the model, performance on Center III data remained poor**.** This suggests additional cohort-specific factors beyond technical variance.

The poor performance of the models on Center III data may be attributed to its small sample size and imbalanced patient characteristics. Notably, only 7 of 19 patients in Center III had a KPS of 90-100%, compared to 66 of 83 and 61 of 77 patients in Centers I and II, respectively. Given that KPS is a strong prognostic factor, its underrepresentation may have compromised model performance. Moreover, advanced metastatic burden was rare in Center III, with only 4 patients presenting with four or more brain metastases. These cohort differences likely limited the generalizability of models trained on the more diverse populations from the other centers.

Taken together, these findings highlight the complexity of building robust, generalizable radiomic models across institutions**.** Even with harmonization efforts and a robust feature elimination pipeline, center-specific differences remain influential. External validation on multiple independent datasets is therefore essential. To support transparency and reproducibility**,** our feature elimination pipeline and trained models (for segmentation and LF prediction) are publicly available. This enables other researchers to test, refine, and compare models, contributing to more generalizable radiomic tools.

In conclusion, analysis of pretherapeutic MRI provides valuable, lesion-specific information for predicting LF in MBM after SRT. Features from the peritumoral margin emerged as the most relevant predictors, potentially reflecting infiltrative behavior not captured by conventional imaging and treatment planning. These findings may help identify high-risk lesions prior to therapy and inform personalized radiotherapy strategies such as expanded margins, dose escalation, or multimodal approaches. Future studies should evaluate the transferability of these markers to other brain metastasis types and further explore strategies to improve model robustness across clinical settings.

## DECLARATIONS

### Funding

This research did not receive any funding

### Ethics approval and consent to participate

The study was approved by the Ethics Committee of the University of Freiburg, Germany (registration number 20-1031), and was conducted in accordance with the Declaration of Helsinki. Additional ethics approvals were obtained at the other participating centers (Kiel: B296/22; Munich: 466/16). Depending on the ethics approvals, the requirement for consent was either waived or, where required, obtained from all included patients for the scientific use of their data.

## References

1. Davies MA, Liu P, McIntyre S, et al. Prognostic factors for survival in melanoma patients with brain metastases. *Cancer*. 2011;117(8):1687-1696. doi:10.1002/cncr.25634

2. Fife KM, Colman MH, Stevens GN, et al. Determinants of Outcome in Melanoma Patients With Cerebral Metastases. *J Clin Oncol*. 2004;22(7):1293-1300. doi:10.1200/JCO.2004.08.140

3. Schadendorf D, van Akkooi ACJ, Berking C, et al. Melanoma. *The Lancet*. 2018;392(10151):971-984. doi:10.1016/S0140-6736(18)31559-9

4. Neal MT, Chan MD, Lucas JT, et al. Predictors of Survival, Neurologic Death, Local Failure, and Distant Failure After Gamma Knife Radiosurgery for Melanoma Brain Metastases. *World Neurosurg*. 2014;82(6):1250-1255. doi:10.1016/j.wneu.2013.02.025

5. Reese RA, Lamba N, Catalano PJ, Cagney DN, Wen PY, Aizer AA. Incidence and Predictors of Neurologic Death in Patients with Brain Metastases. *World Neurosurg*. 2022;162:e401-e415. doi:10.1016/j.wneu.2022.03.028

6. Eigentler TK, Figl A, Krex D, et al. Number of metastases, serum lactate dehydrogenase level, and type of treatment are prognostic factors in patients with brain metastases of malignant melanoma. *Cancer*. 2011;117(8):1697-1703. doi:10.1002/cncr.25631

7. Raizer JJ, Hwu WJ, Panageas KS, et al. Brain and leptomeningeal metastases from cutaneous melanoma: Survival outcomes based on clinical features. *Neuro-Oncol*. 2008;10(2):199-207. doi:10.1215/15228517-2007-058

8. Hasegawa T, Kondziolka D, Flickinger JC, Germanwala A, Lunsford LD. Brain Metastases Treated with Radiosurgery Alone: An Alternative to Whole Brain Radiotherapy? *Neurosurgery*. 2003;52(6):1318-1326. doi:10.1227/01.NEU.0000064569.18914.DE

9. Ghia AJ, Tward JD, Anker CJ, Boucher KM, Jensen RL, Shrieve DC. Radiosurgery for melanoma brain metastases: the impact of hemorrhage on local control. *J Radiosurgery SBRT*. 2014;3(1):43-50.

10. Ramakrishna N, Margolin KA. Multidisciplinary Approach to Brain Metastasis from Melanoma; Local Therapies for Central Nervous System Metastases. *Am Soc Clin Oncol Educ Book*. 2013;(33):399-403. doi:10.14694/EdBook_AM.2013.33.399

11. Yamamoto M, Higuchi Y, Sato Y, Aiyama H, Kasuya H, Barfod BE. Stereotactic Radiosurgery for Patients with 10 or More Brain Metastases. *Prog Neurol Surg*. 2019;34:110-124. doi:10.1159/000493056

12. Gondi V, Pugh SL, Tome WA, et al. Preservation of Memory With Conformal Avoidance of the Hippocampal Neural Stem-Cell Compartment During Whole-Brain Radiotherapy for Brain Metastases (RTOG 0933): A Phase II Multi-Institutional Trial. *J Clin Oncol*. 2014;32(34):3810-3816. doi:10.1200/JCO.2014.57.2909

13. Sperduto PW, Jiang W, Brown PD, et al. Estimating Survival in Melanoma Patients With Brain Metastases: An Update of the Graded Prognostic Assessment for Melanoma Using Molecular Markers (Melanoma-molGPA). *Int J Radiat Oncol*. 2017;99(4):812-816. doi:10.1016/j.ijrobp.2017.06.2454

14. Bian SX, Routman D, Liu J, et al. Prognostic factors for melanoma brain metastases treated with stereotactic radiosurgery. *J Neurosurg*. 2016;125(Supplement_1):31-39. doi:10.3171/2016.8.GKS161359

15. Rizzo S, Botta F, Raimondi S, et al. Radiomics: the facts and the challenges of image analysis. *Eur Radiol Exp*. 2018;2(1). doi:10.1186/s41747-018-0068-z

16. Bhatia A, Birger M, Veeraraghavan H, et al. MRI radiomic features are associated with survival in melanoma brain metastases treated with immune checkpoint inhibitors. *Neuro-Oncol*. 2019;21(12):1578-1586. doi:10.1093/neuonc/noz141

17. Habibi MA, Rashidi F, Habibzadeh A, Mehrtabar E, Arshadi MR, Mirjani MS. Prediction of the treatment response and local failure of patients with brain metastasis treated with stereotactic radiosurgery using machine learning: A systematic review and meta-analysis. *Neurosurg Rev*. 2024;47(1):199. doi:10.1007/s10143-024-02391-3

18. Grosu AL, Frings L, Bentsalo I, et al. Whole-brain irradiation with hippocampal sparing and dose escalation on metastases: neurocognitive testing and biological imaging (HIPPORAD) – a phase II prospective randomized multicenter trial (NOA-14, ARO 2015–3, DKTK-ROG). *BMC Cancer*. 2020;20(1):532. doi:10.1186/s12885-020-07011-z

19. Lin NU, Lee EQ, Aoyama H, et al. Response assessment criteria for brain metastases: proposal from the RANO group. *Lancet Oncol*. 2015;16(6):e270-278. doi:10.1016/S1470-2045(15)70057-4

20. Carré A, Klausner G, Edjlali M, et al. Standardization of brain MR images across machines and protocols: bridging the gap for MRI-based radiomics. *Sci Rep*. 2020;10(1):12340. doi:10.1038/s41598-020-69298-z

21. Avants B, Tustison NJ, Song G. Advanced Normalization Tools: V1.0. *Insight J*. Published online July 29, 2009. doi:10.54294/uvnhin

22. Smith SM. Fast robust automated brain extraction. *Hum Brain Mapp*. 2002;17(3):143-155. doi:10.1002/hbm.10062

23. Isensee F, Jaeger PF, Kohl SAA, Petersen J, Maier-Hein KH. nnU-Net: a self-configuring method for deep learning-based biomedical image segmentation. *Nat Methods*. 2021;18(2):203-211. doi:10.1038/s41592-020-01008-z

24. Karami E, Soliman H, Ruschin M, et al. Quantitative MRI Biomarkers of Stereotactic Radiotherapy Outcome in Brain Metastasis. *Sci Rep*. 2019;9(1):19830. doi:10.1038/s41598-019-56185-5

25. van Griethuysen JJM, Fedorov A, Parmar C, et al. Computational Radiomics System to Decode the Radiographic Phenotype. *Cancer Res*. 2017;77(21):e104-e107. doi:10.1158/0008-5472.CAN-17-0339

26. Bradshaw TJ, Huemann Z, Hu J, Rahmim A. A Guide to Cross-Validation for Artificial Intelligence in Medical Imaging. *Radiol Artif Intell*. 2023;5(4):e220232. doi:10.1148/ryai.220232

27. Cox DR. Regression models and life-tables. *J R Stat Soc Ser B Methodol*. 1972;34(2):187-202.

28. Chang E, Joel MZ, Chang HY, et al. Comparison of radiomic feature aggregation methods for patients with multiple tumors. *Sci Rep*. 2021;11(1):9758. doi:10.1038/s41598-021-89114-6

29. Stera S, Balermpas P, Blanck O, et al. Stereotactic radiosurgery combined with immune checkpoint inhibitors or kinase inhibitors for patients with multiple brain metastases of malignant melanoma. *Melanoma Res*. 2019;29(2):187-195. doi:10.1097/CMR.0000000000000542

30. Siam L, Bleckmann A, Chaung HN, et al. The metastatic infiltration at the metastasis/brain parenchyma-interface is very heterogeneous and has a significant impact on survival in a prospective study. *Oncotarget*. 2015;6(30):29254-29267. doi:10.18632/oncotarget.4201

31. Schatlo B, Stockhammer F, Barrantes-Freer A, et al. 5-Aminolevulinic Acid Fluorescence Indicates Perilesional Brain Infiltration in Brain Metastases. *World Neurosurg X*. 2020;5:100069. doi:10.1016/j.wnsx.2019.100069

32. Fiss I, Hussein A, Barrantes-Freer A, et al. Cerebral metastases: do size, peritumoral edema, or multiplicity predict infiltration into brain parenchyma? *Acta Neurochir (Wien)*. 2019;161(5):1037-1045. doi:10.1007/s00701-019-03842-3

33. Shofty B, Artzi M, Shtrozberg S, et al. Virtual biopsy using MRI radiomics for prediction of BRAF status in melanoma brain metastasis. *Sci Rep*. 2020;10(1):6623. doi:10.1038/s41598-020-63821-y

34. Su CQ, Chen XT, Duan SF, et al. A radiomics-based model to differentiate glioblastoma from solitary brain metastases. *Clin Radiol*. 2021;76(8):629.e11-629.e18. doi:10.1016/j.crad.2021.04.012

35. Mi H, Yuan M, Suo S, et al. Impact of different scanners and acquisition parameters on robustness of MR radiomics features based on women’s cervix. *Sci Rep*. 2020;10(1):20407. doi:10.1038/s41598-020-76989-0

## Tables

| **Table 1 -** Clinical parameters, per series. *Abbreviations*: MBM = melanoma brain metastases; KPS = Karnofsky performance scale; NA: not available; RT = radiotherapy. * = Systemic therapy was defined as chemotherapy, immunotherapy or targeted therapy before (>14 days before), during (within 14 days before or after RT) and after (>14 days after) radiotherapy. Melanoma Molecular-Graded Prognostic Assessment Score was assessed at MBM diagnosis[13]. | | | | |
|---|---|---|---|---|
| Institution | Total (n=179) | Center I (n=83) | Center II (n=77) | Center III (n=19) |
| Number of lesions | 517 | 247 | 226 | 44 |
| Sex (female/male) | 63/116 | 28/55 | 27/50 | 8/11 |
| Age at RT start (mean ± SD) | 61.9 ±14.2 Y | 59.9 ±14.6 Y | 63.9 ±13.6 Y | 62.9 ±14.7 Y |
| KPS | | | | |
| 90-100% | 134 | 66 | 61 | 7 |
| 70-80% | 38 | 16 | 13 | 9 |
| 60% | 7 | 1 | 3 | 3 |
| Number of MBM (series) | | | | |
| 1 | 55 | 22 | 26 | 7 |
| 2-3 | 54 | 20 | 26 | 8 |
| 4 or more | 70 | 41 | 25 | 4 |
| Melanoma Molecular-Graded Prognostic Assessment Score | | | | |
| 0-1 | 26 | 19 | 7 | NA |
| 1.5-2 | 74 | 43 | 31 | NA |
| 2.5-3 | 41 | 17 | 24 | NA |
| 3.5-4 | 19 | 4 | 15 | NA |
| Molecular target | | | | |
| BRAF-mutation | 87 | 47 | 31 | 9 |
| NRAS | 23 | 14 | 8 | 1 |
| no driver mutation | 69 | 22 | 38 | 9 |
| Systemic therapy* | | | | |
| before RT (yes/no/NA) | 116/44/21 | 44/40/0 | 64/4/10 | 8/0/11 |
| during RT (yes/no/NA ) | 113/47/21 | 42/41/1 | 63/6/13 | 12/0/7 |
| after RT (yes/no/NA) | 136/27/18 | 53/27/4 | 65/0/13 | 18/0/1 |
| Immunotherapy | | | | |
| before RT (yes/no/NA) | 81/91/9 | 23/61/0 | 50/19/9 | 8/11/0 |
| during RT (yes/no/NA) | 86/84/11 | 26/58/0 | 48/19/11 | 12/7/0 |
| after RT (yes/no/NA) | 112/57/12 | 40/44/0 | 55/11/12 | 17/2/0 |

**Table 2** – Radiotherapy specifications per lesion. *Abbreviations*: BED = biologically effective dose, SRT = stereotactic radiotherapy, SRS = stereotactic radiosurgery, SFRT = stereotactic fractionated radiotherapy

| | Total number of lesions (n=517) | Center I (n=247) | Center II (n=226) | Center III (n=44) |
|---|---|---|---|---|
| SRS | 349 | 93 | 218 | 38 |
| SFRT 12-14 fractions | 147 | 147 | 0 | 0 |
| SFRT 2-4 fractions | 21 | 7 | 8 | 6 |
| SRS dose (median, range) | 20 (16-20) Gy | 20 (18-20) Gy | 18 (16-20) Gy | 20 (18-20) Gy |
| SFRT dose (median, range) | 51 (24-51) Gy à 4.25 (3-8) Gy | 51 (35-51) Gy à 4.25 (3-5) Gy | 24 Gy à 8 Gy | 35 Gy à 5 Gy |
| BED α/ß=2 Gy (median, range) | 180 (97.5-220) Gy | 159.4 (97.5-220) Gy | 180 (120-220) Gy | 220 (122.5-220) Gy |
| BED α/ß=10 Gy (median, range) | 60 (41.6-72.7) Gy | 72.7 (50.4-72.7) Gy | 50.4 (41.6-60) Gy | 60 (50.4-60) Gy |
| Number of fractions (median, range) | 1 (1-14) | 12 (1-14) | 1 (1-3) | 1 (1-7) |